\pdfoutput=1
\documentclass[review]{elsarticle}
\usepackage[a4paper,left=2.5cm,right=2.5cm,top=2.5cm,bottom=2.5cm]{geometry}
\usepackage[utf8]{inputenc}
\usepackage[T1]{fontenc}
\usepackage[mathcal]{eucal}
\usepackage{amsfonts, amsmath, amssymb, bm, mathrsfs}
\usepackage{graphicx, epsfig, color, colortbl, subfigure, adjustbox}
\usepackage{array, import, float, multirow}
\usepackage{relsize, lscape, rotating}
\usepackage{natbib}                   
\usepackage{ulem, soul, lmodern}
\usepackage[hidelinks]{hyperref}

\hypersetup{ pdftitle  = {Continuum modeling of the effect of surface area growth due to crushing and damage on the permeability of granular rocks},
             pdfauthor = {S. Esna Ashari} }

\biboptions{sort&compress}	
\journal{Acta Geotechnica}
\begin{document}
\begin{frontmatter}
\title{\textbf{Continuum modeling of the effect of surface area growth due to crushing and damage on the permeability of granular rocks}}
\author[NWU]{S. Esna Ashari}
\ead{s-esnaashari@u.northwestern.edu}
\author[IITK]{A. Das}
\ead{arghya@iitk.ac.in}
\author[NWU]{G. Buscarnera}
\ead{g-buscarnera@northwestern.edu}
\cortext[correspondingauthor]{Corresponding author}
\address[NWU]{Northwestern University, Department of Civil and Environmental Engineering, USA}
\address[IITK]{Indian Institute of Technology Kanpur, Department of Civil Engineering, India}
%
%
%
%
%
\begin{abstract}
This paper discusses a continuum approach to track the evolution of permeability in granular rocks by accounting for the combined effect of porosity changes, grain breakage and cement bond damage. To account for such a broad range of microscopic processes under general loading paths, the Breakage Mechanics theory is used and the computed mechanical response is linked with the Kozeny equation, i.e. a permeability model able to evaluate the reduction of the hydraulic conductivity resulting from the simultaneous loss of porosity and growth of surface area. In particular, the evolution of the internal variables of the model has been linked to idealized geometric schemes at particle scale, with the goal to distinguish the contribution of the fines generated by the disaggregation of the cement matrix from that of the broken fragments resulting from the crushing of the skeleton. Compression/flow experiments available in the literature for different granular rocks are used to validate the proposed methodology. The analyses illustrate that the drop of the permeability of damaged rocks would be severely underestimated without an accurate computation of the growth of surface area, as well as that the distributed fragmentation of skeleton particles tends to have stronger implications than the generation of cement fines. These findings, along with the satisfactory agreement between model predictions and experiments, stress the benefits of adopting microstructure-based constitutive laws for the analysis of coupled hydro-mechanical problems.
\end{abstract}
\end{frontmatter}
%
%
%
\section{Introduction}
Fluid flow in reservoir rocks is invariably coupled with various deformation mechanisms associated with grain breakage, compaction, strain localization and fracture. The interplay between such microstructural alterations of the solid matrix and pore-scale flow controls the overall permeability of rocks. Understanding such relations is pivotal for the success of any activity involving injection/extraction of fluids, such as oil and gas extraction and underground storage of by-products \cite{Hettema_2000, Gens_2010, Trebotich_2014}. For granular porous media, the permeability is a function of the grain size distribution, as well as of how the degree of particle sorting evolves during deformation \cite{Freeze_1977, Odong_2007}. Such distributed fragmentation of the skeletal grains is often accompanied by the propagation of micro-cracks across the cement matrix bonding the grains, which in turn represents a further cause of surface area growth \cite{Hu_2007, Motamedi_2016}.\\
The variation of permeability in rocks has been widely studied from an experimental standpoint  \cite{Katz_1986, Zhou_1997, Ostermeier_2001, Liu_2004} and numerous empirical correlations were proposed \cite{Sahimi_1993, Sahimi_1995}. Similarly, a wide range of numerical  \cite{Zijl_2001, Chen_2003} and analytical studies \cite{Acuna_1995, Karacan_2003} have been dedicated to this topic. The physical interaction between fluid flow and deformation processes; i.e. hydro-mechanical (HM) coupling, in rocks has also attracted considerable attention \cite{Olsson_2001, Rutqvist_2003}; especially when flow-deformation analyses can be dramatically affected by the porosity-dependence of rock permeability. In this context, usual HM continuum formulations for granular solids tend to lack coordinated mechanical and hydraulic constitutive laws based on the same microstructural descriptors, and often resort to empirical permeability functions only coupled with porosity changes. Several models, however, nowadays can take into account the evolution of the material microstructure during inelastic deformation, thus providing new avenues to account for the important effects of the changes in specific surface area that may result from micro-fracturing and grain crushing.\\
This paper aims to discuss a continuum modeling approach that quantifies changes in granular rocks permeability by tracking the evolution of microstructural descriptors linked with the simultaneous loss of porosity and growth of surface area. For this purpose, a continuum model originally developed based on the Breakage Mechanics theory \cite{einav_2007_a, einav_2007_b, tengattini_2014} is used to track the progression of microscopic grain fragmentation and cement disaggregation through two dedicated internal variables, a Breakage index $B$ and a Damage index $D$, linked with the grain size distribution and cement size distribution, respectively. Such micromechanical features, together with the ability to track changes in porosity and surface area, make the Breakage Mechanics theory suited for hydro-mechanical analyses at the constitutive level \cite{Buscarnera_2012_1, Zhang_2014, Zhang_2015}. Former studies have indeed demonstrated its applicability to the study of permeability evolution in rocks by linking different classes of Breakage models to a Kozeny law describing the hydraulic conductivity of granular rocks \cite{Nguyen_2009, EsnaAshari_2016}. In the following, the latter idea is further explored for the case of unconsolidated granular rocks (Glass Beads and Ottawa sand) and weakly cemented granular rocks (Otter Sherwood sandstone), with the goal to evaluate the performance of the Breakage Mechanics model against the available experimental data. 
%
\section{Technical background}\label{Technical_background}
\subsection{Permeability models for granular materials}\label{Permeability_models}
\begin{figure}[!ht]
\centering
\includegraphics[width=7.0cm]{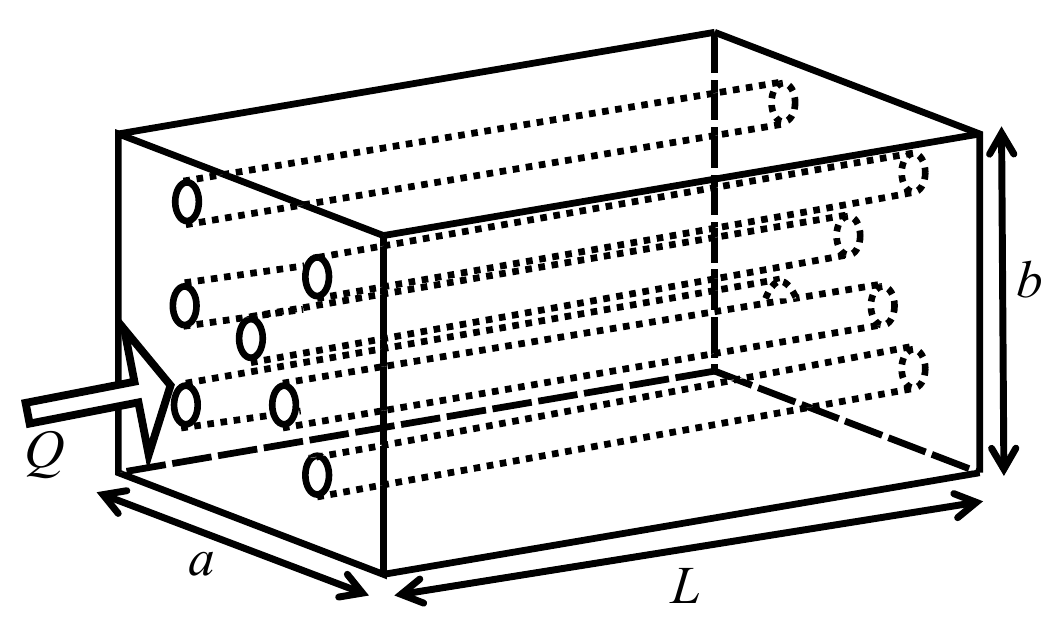}
\caption{ Capillary tube model.}\label{figure1}
\end{figure}

Numerous empirical formulae, capillary models, statistical models and hydraulic radius theories have been proposed to estimate the permeability of granular rocks. Capillary tube models are among the simplest and most widely used models from which many other existing permeability laws are formulated, in that they provide a mechanistic explanation for the Darcy's law controlling the macroscopic flow of a viscous fluid through a porous medium. In a one-dimensional context such laws can be written as follows: 
\begin{equation}\label{darcy}
Q = \frac{-\kappa A }{\mu} \frac{\partial {P}}{\partial x}
\end{equation}
where $Q$ is the flowrate of the formation, $\kappa$ is the intrinsic permeability of the medium, $A$ is the cross-sectional area available for flow, $\mu$ is the viscosity of the fluid and $\frac{\partial {P}}{\partial x}$ is the pressure gradient. The negative sign in Equation (\ref{darcy}) indicates that the fluid flows from high pressure to low pressure.\\
The capillary tubes model is formulated from the Hagen-Poisseuille's law which governs the steady flow in a single straight circular capillary tube of diameter $d$ \cite{bear_1972}:
\begin{equation}\label{Poisseuille}
Q = \frac{-\pi d^4 \rho g}{128\mu} \frac{\partial \varphi}{\partial x}
\end{equation}
where $\rho$ is the fluid density, $g$ is the gravitational acceleration and $\varphi$ is the pressure head. The analogy between Equation (\ref{darcy}) and Equation (\ref{Poisseuille}) is apparent considering that $\varphi=P/\rho g$ for static pressure, while the medium permeability $\kappa$ equals $d^2/32$. The capillary tubes model considers a bundle of such parallel tubes, all of the same diameter $d$ similar to Figure \ref{figure1} and gives the specific discharge $\bar{q}$ through the porous block for the case of $N$ tubes per unit area of cross-section of the model as:
\begin{equation}\label{Tube}
\bar{q}=Q/ab = -{N} \frac{\pi d^4 \rho g}{128\mu} \frac{\partial \varphi}{\partial x}
\end{equation}
Since the porosity $n$ of this model is defined as $n=N(\pi d^2/4)L/L$, Equation (\ref{Tube}) can be rewritten as:
\begin{equation}\label{Tube_porosity}
\bar{q}=- \frac{\kappa \rho g}{\mu} \frac{\partial \varphi}{\partial x};\;\;\kappa = n d^2/32
\end{equation}
which is again analogous to Darcy's law.\\
These principles are used in the Kozeny model \cite{kozeny_1927}, in that the porous medium is treated as a bundle of capillary tubes of equal length. The cross-section of the tubes is not necessarily circular in this model, but similarly to the original capillary tubes model only the fluid velocity component oriented along the tube direction is taken into account in the model derivation. By expressing $\bar{q}$ with respect to a unit volume of solid rather than a unit volume of the porous medium, the Kozeny model derives:
\begin{equation}\label{kozeny}
\bar{q}=- \frac{\kappa}{\mu} \nabla{P};\;\;\kappa = \frac{c_0 n^3}{S^2}
\end{equation}
where $c_0$ is the Kozeny's constant (e.g., $c_0=0.5$ for circular channels) and $S$ is the specific surface area (i.e. the total surface area of solid material per unit of volume) which is inversely proportional to the hydraulic radius $R$ of the porous medium (i.e. the ratio between the cross-sectional area and the wetted perimeter) as $S = n/R$. The expression of $\kappa$ in Equation (\ref{kozeny}) is referred to as Kozeny's equation and has been used widely by many researchers to evaluate the permeability of granular media, and it will be used hereafter to assess the evolution of granular rock permeability in presence of simultaneous changes of porosity and surface area. .\\
\subsection{Breakage Mechanics models}\label{Breakage_models}
\begin{figure}[!ht]
\centering
\includegraphics[width=7.0cm]{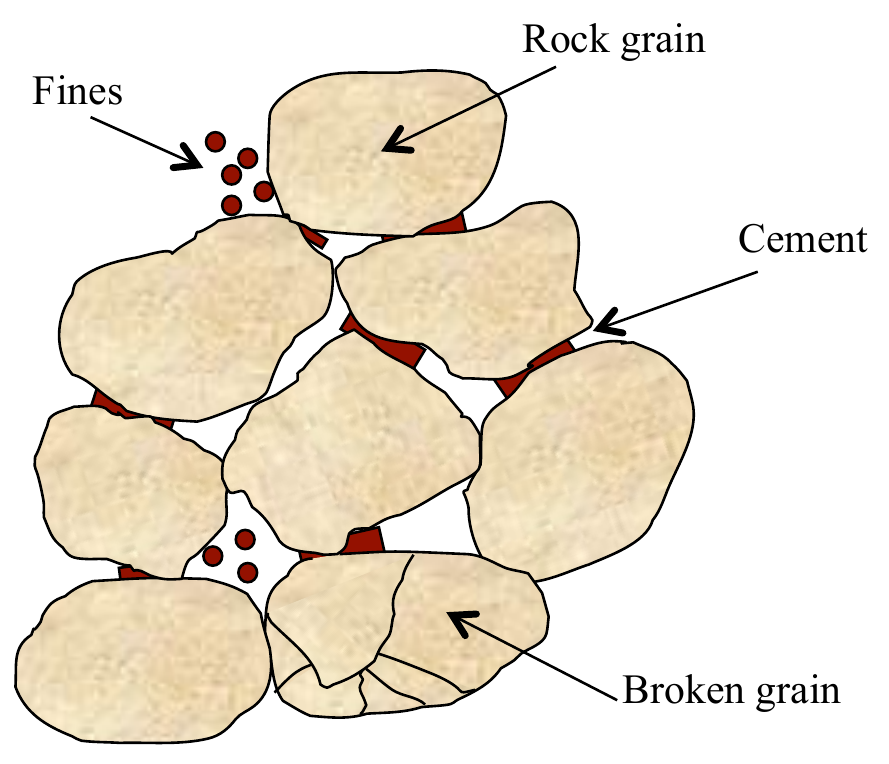}
\caption{Broken cement between grains and the generated pseudo grains (fines).}\label{figure2}
\end{figure}

\begin{figure}[!ht]
\centering
\includegraphics[width=13.0cm]{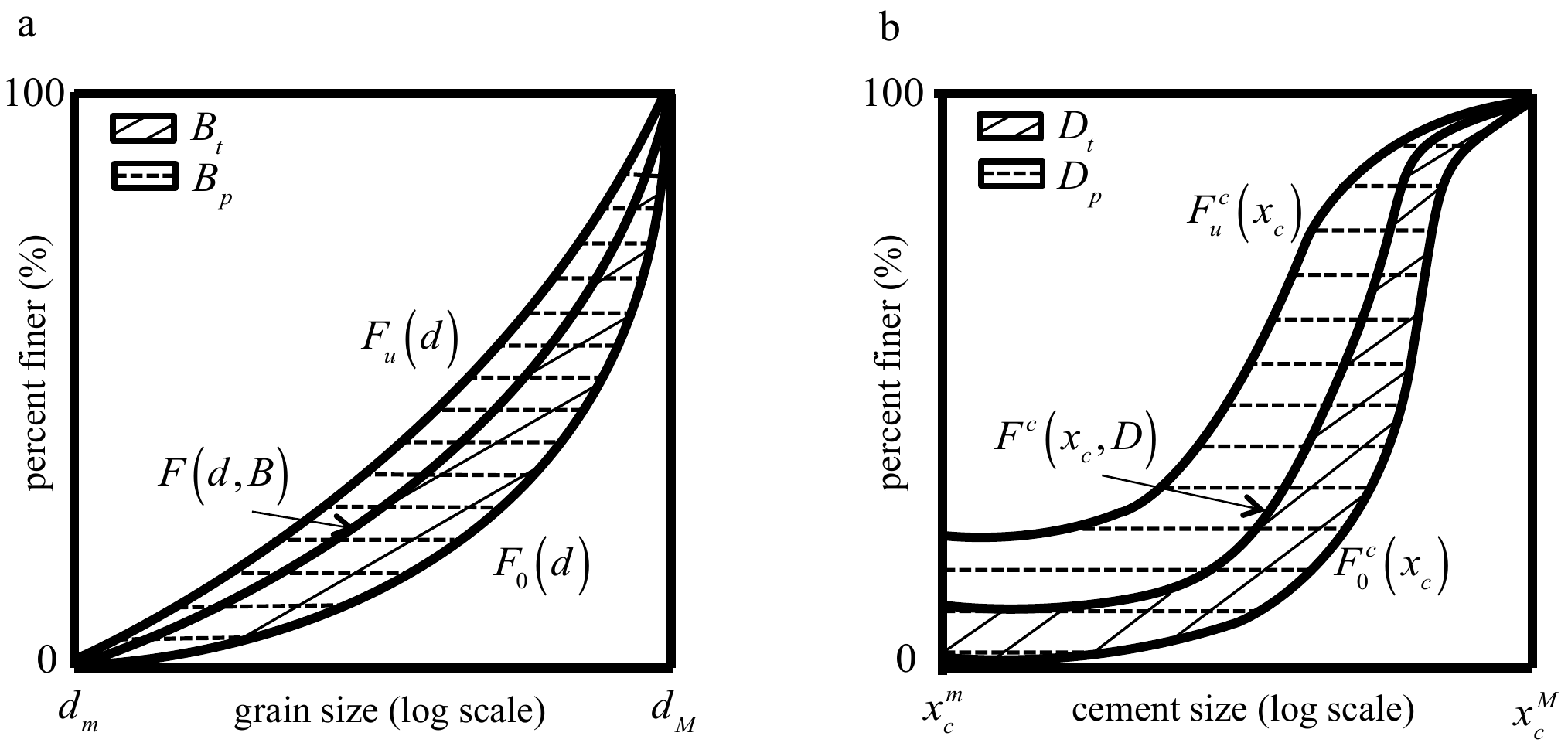}
\caption{Grain size distribution evolution (a) Granular phase for $B$ calculation. (b) Cement phase for $D$ calculation.}\label{figure3}
\end{figure}

Modeling the deformation of geomaterials in a continuum mechanics framework implies challenges associated with the representation of the macroscopic effects of local inter-particle interactions. Within this context, crushing phenomena are characterized by the further challenge of simulating the implication of an evolving particle gradation.\\
The Breakage Mechanics framework \cite{einav_2007_a,einav_2007_b} is an example of continuum theory for crushable granular materials in which the particle-continuum duality is captured by combining thermodynamics principles with micromechanical arguments. Specifically, the theory is capable of describing the evolution of the grain size distribution (gsd) by means of a breakage internal variable associated with the release of elastic strain energy due to local grain fracture. This feature creates an explicit link between micro-scale mechanisms and macro-scale mechanical response, thus offering the significant advantage of tracking microstructural attributes in addition to usual global measures of deformation.\\
This framework has been recently augmented by  \citet{tengattini_2014} to cope with cemented granular systems. Specifically, an internal variable reflecting the development of micro-cracks in the cement phase was added to the formulation, with the goal to mimic the implications of the progressive rupture of the bonds bridging the particles (Figure \ref{figure2}). This has led to an enhanced Breakage-Damage formalism particularly convenient to track surface area changes in lightly cemented granular materials, where the damage of the cement interacts with the fragmentation of the skeletal particles.\\
Hereafter the components of the Breakage Mechanics formulations essential for this work are briefly discussed, while a more detailed description can be found in \citet{einav_2007_a} and \citet{tengattini_2014}.\\
%
According to the Breakage Mechanics theory, gsd can be used to define the current state of a crushable granular material and it can be tracked by means of a scalar internal variable named Breakage $B$, defined as: $B=B_t/B_p$. $B_t$ and $B_p$ can be deduced by using the initial and ultimate cumulative gsd functions $F_0(d)$ and $F_u(d)$ similar to Figure \ref{figure3}a where $d$ represents the grain size. While the initial gsd is typically measured to define the initial state of the material, the ultimate gsd is often assumed to be represented by power-law distribution of fractal nature, similar to Equation (\ref{fractal}) with a fractal dimension $\alpha$ \cite{turcotte_1986,McDOWELL_1996}.
\begin{equation}\label{fractal}
F_u(d)=\frac{d^{3-\alpha}-d_m^{3-\alpha}}{d_M^{3-\alpha}-d_m^{3-\alpha}}
\end{equation}
whre $d_m$ and $d_M$ are the minimum and maximum grain size at the ultimate gsd and the fractal dimension $\alpha$ is often chosen in the range 2.5-2.9. The current cumulative gsd can be expressed as a function of $B$ as follows: 
\begin{equation}\label{current_cpdf}
F(d,B) = (1-B)F_0(d)+BF_u(d)
\end{equation}
For cemented granular materials, micro-cracking of the cement phase is associated with a decrease in the size of the mechanically active inter-granular bond, which results into cement bridges of smaller cross-sectional area. Using this assumption, it is possible to define an effective cement size ($x_c$) distribution, which describes the evolution of the damage in the cement phase. Similar to breakage $B$, another scalar variable, i.e. damage $D$, is defined in the Breakage-Damage theory as: $D=D_t/D_p$. Again $D_t$ and $D_p$ can be determined by using the initial and ultimate cumulative effective cement size distribution functions $F_0^c(x_c)$ and $F_u^c(x_c)$ similar to Figure \ref{figure3}b. The initial effective cement size distribution can be calculated from X-Ray tomographic imaging of the material \cite{David_1998} while the ultimate effective cement size distribution can be hypothesized as a constant function implying the theoretical final state is a completely uncemented granular system; i.e., $D = 1$.\\
The current cumulative effective cement size distribution can be defined as
\begin{equation}\label{current_cpdf_cement}
F^c(x_c,D) = (1-D)F_0^c(x_c)+DF_u^c(x_c)
\end{equation}
%
The evolution of such two internal variables is linked with the storage and dissipation of energy, which implies the repartition of the work input done at the boundaries of an RVE, $\tilde{W}$, into the sum of the rate of stored energy (which under isothermal conditions can be quantified through the Helmholtz Free Energy, $\dot{ \Psi}$) and the rate of energy dissipation, $\tilde{\Phi}$; $\tilde{W}=\dot{ \Psi}+\tilde{\Phi}$ \cite{tengattini_2014}. The Helmholtz Free Energy in a system with multiple interacting solid phases can be expressed as the volume-average of strain energy contributions associated with the cement and granular phases, as follows:
\begin{equation}\label{Helmholtz}
\Psi = \Psi^g\varphi^g+\Psi^c\varphi^c
\end{equation}
where $\varphi^g$ and $\varphi^c$ are the the volume fractions of grain and cement ($\varphi^g+\varphi^c=1$). $\Psi^g$ and $\Psi^c$ are the average stored energy within the grain and cement volume fractions, respectively, which can be written as functions of the elastic strain energy in an unbroken/undamaged reference system ($\Psi_r^g$ for the grain and $\Psi_r^c$ for the cement phase):   
\begin{equation}\label{flow_rules}
\begin{split}
&\Psi^g=\Psi_r^g(\varepsilon_{ij}^e)(1-\vartheta^g B)\\ 
&\Psi^c=\Psi_r^c(\varepsilon_{ij}^e)(1-D)\\
\end{split}
\end{equation}
where $\vartheta^g$ is a grading index emerging from a statistical homogenization procedure based on initial and ultimate grain size distributions \cite{einav_2007_a}. The dependence of the $\Psi$ function on breakage and damage indices emerges from statistical homogenization techniques, according to which the change in elastic properties is linked to the release of strain energy associated with permanent changes of the microstructural attributes. The reference energy (for both grain and cement) can be subdivided into two distinct parts: the volumetric part $\Psi_r^{v,g}, \Psi_r^{v,c}$ and the shear part $\Psi_r^{s,g}, \Psi_r^{s,c}$; i.e. $\Psi_r^g=\Psi_r^{v,g}+\Psi_r^{s,g}$ and $\Psi_r^c=\Psi_r^{v,c}+\Psi_r^{s,c}$, and each of these reference elastic potentials can be chosen to be either linear or nonlinear (pressure-dependent) functions. Different definitions of these functions as well as the specific forms used in this work are presented in the Appendix.\\
Starting from the Helmholtz Free Energy, the Breakage Energy ($E_B=-\frac{\partial \Psi}{\partial B}$) and the Damage Energy ($E_D=-\frac{\partial \Psi}{\partial D}$) can be computed, which represent the strain energy released by the material during the breakage of the granular skeleton or the fragmentation of the cement bonds, respectively. Finally, by employing a coupled dissipation function based on the simultaneous presence of contributions due to breakage, damage and cohesive-frictional processes \cite{einav_2007_b,tengattini_2014}, it is possible to derive the following yielding condition:
\begin{equation}\label{yield}
y = \frac{(1-B)^2E_B}{E_{BC}}+\frac{(1-D)^2E_D}{E_{DC}}+\frac{q^2}{(Mp+c(1-D))^2}-1\leq 0
\end{equation}
where $M$ is the critical state parameter which is a function of the friction angle $\phi$; i.e. $M=6\sin \phi/(3-\sin\phi)$, $c$ is the cohesion, $p=\frac{\partial \Psi}{\partial \varepsilon_v^e}$ is the mean stress and $q=\frac{\partial \Psi}{\partial \varepsilon_s^e}$ is the differential stress. $E_{BC}$ and $E_{DC}$ are the critical breakage and damage energies; i.e. the energy thresholds at which the respective phases initiate the inelastic processes, in other words, material parameters to be calibrated.\\
\subsection{Connection between the permeability model and the Breakage Mechanics models}\label{permeability_breakage}
As it can be seen in Equation (\ref{kozeny}), the permeability is a function of porosity, as well as of the surface area of the solid. Therefore for geomaterials exhibiting extensive micro-fracturing due to grain crushing and cement damage, the permeability tends to evolve not only because of porosity changes, but also because of alterations of their initial surface area \cite{EVANS_1997}. While most constitutive models for granular materials do not incorporate the effects of an evolving microstructure into permeability predictions \cite{SHELDON_2006}, breakage models can circumvent this problem  by linking the evolution of the surface area to their internal variables $B$ and $D$ \cite{nguyen_2012_1}.\\
This logic can be illustrated in case of a simple assembly of spherical particles, for which the specific surface area $S$ can be written in terms of the harmonic mean grain size $D_{H}$. A specific grain fraction of the assembly, the surface area scales inversely with the hydraulic diameter, such as: $S=6/D_H$ \cite{bear_1972}. Hence, for a polydisperse material characterized by a probability density distribution $f(d)=\frac{\operatorname{d}F(d)}{\operatorname{d}d}$ for the grain size, the hydraulic diameter can be computed as the harmonic mean across all the fractions in the assembly, as follows:
\begin{equation}\label{harmonic_mean}
\frac{1}{D_H}= \int_{d_m}^{d_M}f(d)\frac{\operatorname{d}d}{d}
\end{equation}
If a fractal distribution is considered for $f(d)$, then $D_H$ will be calculated as:
\begin{equation}\label{harmonic_mean_fractal}
D_H=\frac{2-\alpha}{3-\alpha}\frac{d_M^{3-\alpha}-d_m^{3-\alpha}}{d_M^{2-\alpha}-d_m^{2-\alpha}}
\end{equation}
In the case of a crushable granular material, the evolving $f(d)$ becomes a function of both $B$ and $d$ similar to Equation (\ref{current_cpdf}): $f(d,B) = (1-B)f_0(d)+Bf_u(d)$. By plugging this expression in Equation (\ref{harmonic_mean}), the evolving harmonic mean grain size is derived as:
\begin{equation}\label{harmonic_mean1}
\frac{1}{D_H(B)}= (1-B)\frac{1}{D_{H0}}+B\frac{1}{D_{Hu}}
\end{equation}
where $D_{H0}$ and $D_{Hu}$ are the harmonic mean grain size for the initial and ultimate probability density functions of grains (Equation (\ref{harmonic_mean})). The example above provides an expression of $D_H$ valid for a crushable assembly, and which can be readily computed as a function of the internal variable $B$. In the following, it will be shown that this logic can be extended to the case of cemented granular materials by using data about the accumulation of fines resulting from the disaggregation of a damaged cement matrix to derive an expression similar to Equation (\ref{harmonic_mean1}) in which breakage and damage contributions affect simultaneously the surface area growth.\\

\section{Model simulations}\label{Numerical_analysis}
In this section, the Breakage Mechanics model is calibrated and validated against the hydro-mechanical response of different granular solids. In other words, the model parameters are chosen to capture both the compression response and the evolution of the grain/cement size distributions under different loading paths and thereafter, the model is used to estimate the permeability reduction upon compression; i.e. the ratio between the current permeability $\kappa$ and the initial permeability $\kappa_0$, by applying the Kozeny equation (Equation (\ref{kozeny-permeability-reduction})) as follows:
\begin{equation}\label{kozeny-permeability-reduction}
\frac{\kappa}{\kappa_0}=(\frac{n}{n_0})^3(\frac{D_{H}}{D_{H0}})^2
\end{equation}
where $n$, $n_0$ are the current and initial porosity.

\subsection{Cohesionless granular materials}
\subsubsection{Ottawa sand}
\begin{figure}[!ht]
\centering
\includegraphics[width=16.1cm]{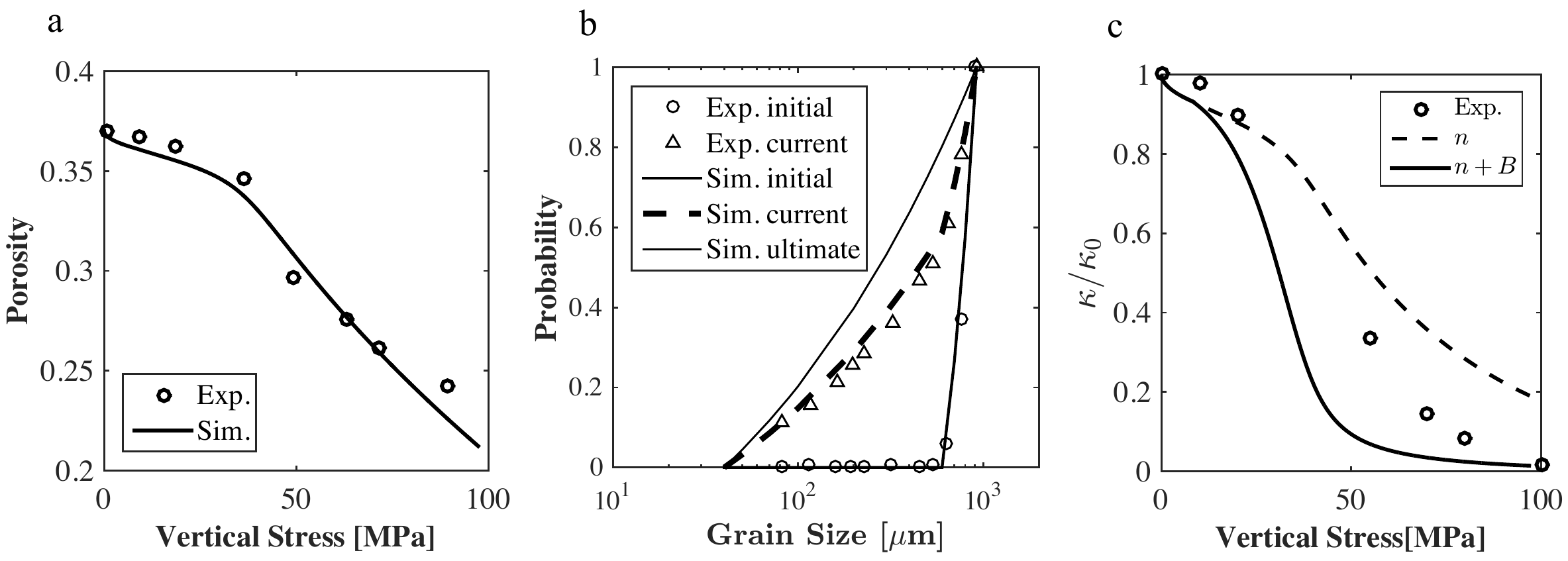}
\caption{(a) Mechanical response (b) Grain size distribution evolution (c) permeability approximation for Ottawa sand under oedometer test [data re-plotted after \cite{DeJong_2009}].}\label{figure4}
\end{figure}

Ottawa sand is the first granular solid studied in this section. To start, the Breakage Mechanics model response under the oedometer stress path is calibrated against the experimental results reported for an initial porosity of $37\%$ in \cite{DeJong_2009}. The calibrated compressive response is plotted in (Figure \ref{figure4}a) in porosity-vertical stress space where $E_{BC}$= 0.35,  $\omega_B$= 68$^{\circ}$, $\overline{K}^g$= 7700 and $\overline{G}^g$ is 4000. $M$ is 1.24 calculated for the friction angle of $\phi$= 31$^{\circ}$ \cite{Cho_2001}.\\
Setting $d_m$= 600 $ \text{$\mu$m}$ and $d_M$= 920 $ \text{$\mu$m}$, the initial cumulative gsd function $F_0(d)$ is approximated by a power law distribution with the power law coefficient of 1, while a fractal distribution with the fractal dimension $\alpha$ of 2.7 with $d_m$= 40 $ \text{$\mu$m}$ and $d_M$= 920 $ \text{$\mu$m}$ is used for ultimate grading $F_u(d)$ and $\vartheta^g$= 0.7 is calculated from $F_0(d)$ and $F_u(d)$. The current grain size distribution is derived based on the final Breakage value at the end of the simulation ($B$=0.73) using Equation (\ref{current_cpdf}). Figure \ref{figure4}b compares the computed gsds with those emerging from the tests, indicating a satisfactory match between model and experimental results.  Most notably, the role of changes in specific surface area is noticed when the permeability loss due to simultaneous compaction and breakage is compared with computations based on porosity changes only (Figure \ref{figure4}c). Neglecting surface area growth implies indeed a severe underestimation of the permeability loss, with predictions that deteriorate as soon as large compressive stresses are attained.\\
Although a good agreement between data and model predictions is found, some discrepancies in the magnitude of permeability drop are noticed, with the model overestimating the permeability decrease compared to the experiment. Such mismatch tends to be larger at intermediate stress levels, and it can be related to the tendency of Equation (\ref{current_cpdf}) to overestimate the amount of breakage-induced grains during the first stages of post-yielding compression due to the assumption of using a linear scaling of the evolving grain size distribution. The mismatch, however, disappears as high compressive stresses are approached, i.e. when the computed gsd curve captures accurately the gsd measured at the end of the experiment.\\
\subsubsection{Glass Beads}
\begin{figure}[!ht]
\centering
\includegraphics[width=16.0cm]{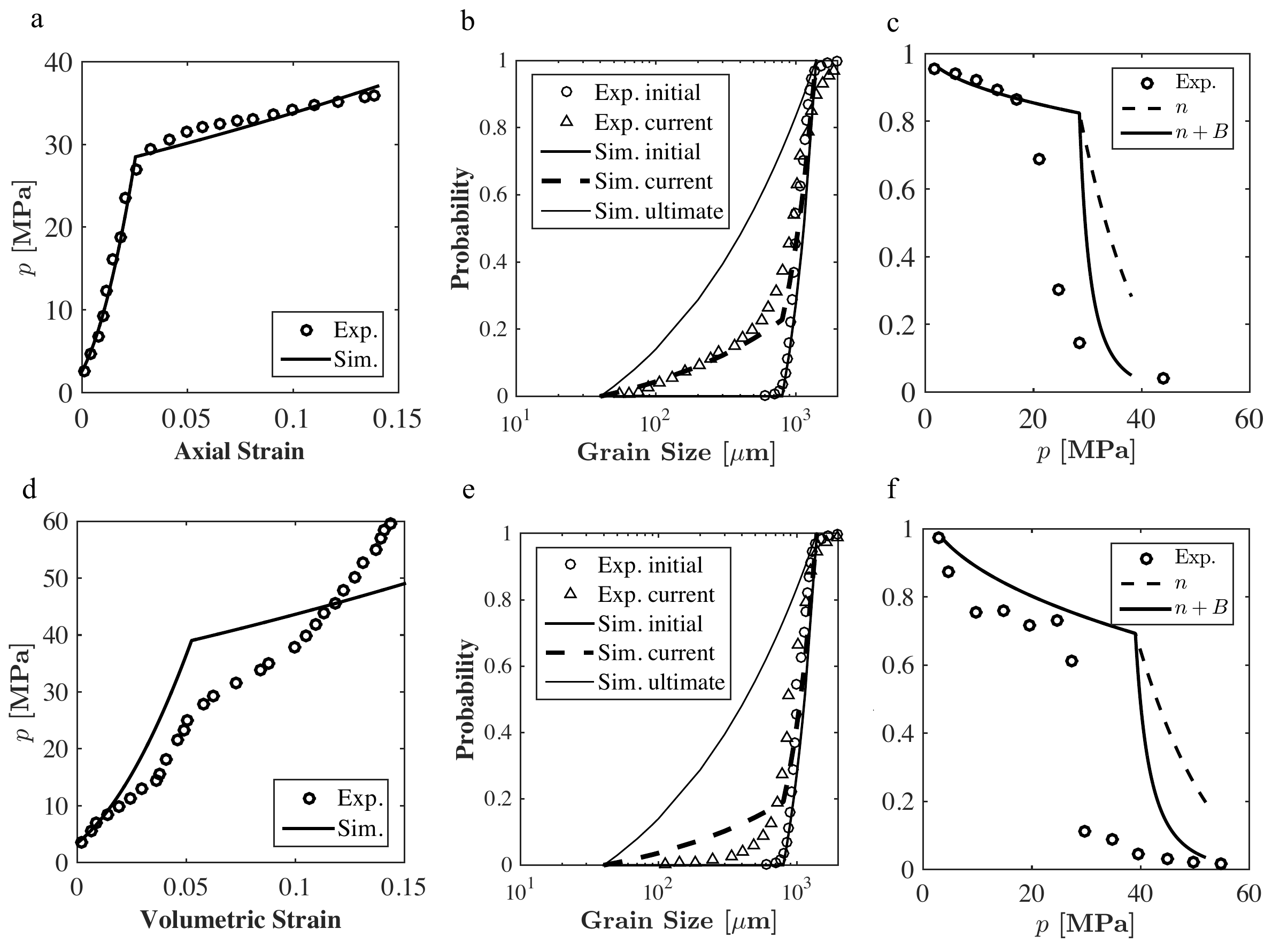}
\caption{(a), (b) and (c) Mechanical response, grain size distribution and permeability approximation for Glass Beads ($r$= 0.6). (d), (e) and (f) Mechanical response prediction, grain size distribution prediction and permeability approximation for Glass Beads ($r$= 1.0) [data re-plotted after \cite{Nguyen_2011}].}\label{figure5}
\end{figure}

As a further example, here the variation of permeability in Glass Beads subjected to compression is discussed. Similar to Ottawa sand, the Breakage Mechanics model was calibrated to capture the mechanical response under the effect of radial loading paths. The latter were imposed by controlling a stress path parameter $r$ defined as the ratio between radial and axial stress increments, $r=\Delta \sigma_r/\Delta \sigma_a$ ($r=1$ representing hydrostatic paths). The experimental data for calibration and validation purposes are derived from previous studies reported by \cite{Nguyen_2011}. Figure \ref{figure5}a illustrates the calibrated response versus the experimental data in the mean pressure $p$-axial strain space for the test $r$= 0.6 with $E_{BC}$= 0.65,  $\omega_B$= 75$^{\circ}$, $\overline{K}^g$= 5300, $\overline{G}^g$= 2000 and $M$= 0.98.
The current gsd is evaluated by assuming $d_m$= 800 $ \text{$\mu$m}$, $d_M$= 1400 $ \text{$\mu$m}$ and power law coefficient of 1.2 for $F_0(d)$ and $d_m$= 40 $ \text{$\mu$m}$, $d_M$= 1400 $ \text{$\mu$m}$ and $\alpha$ of 2.6 for $F_u(d)$ which gives $\vartheta^g$= 0.67. Figure \ref{figure5}b shows that the cumulative gsd calculated by the model for $B$= 0.31 agrees well with the experiment.\\
The permeability reduction for this test is also calculated and reported in Figure \ref{figure5}c. The model and the data mismatch to some extent in terms of the stresses at which the permeability begins to drop. Such a mismatch, however, is not visible in the compression behavior with the model capturing accurately the yield point observed in the test. Such discrepancies can be explained as an outcome of local inelastic processes at the micro-scale taking place prior to yielding, which do not alter significantly the mechanical response of the skeleton but tend to affect more noticeably the hydrologic properties, thus causing a reduction of permeability at slightly smaller stresses than those related with inelastic compaction. While the model does not capture such pre-yielding processes (i.e., it assumes a sharp passage to inelastic behavior) and predicts the initiation of permeability reduction at the same stresses at which the material yields ($p$= 28 MPa in this case), the comparison between data and simulations indicates a good quantitative match of the computed permeability at high stress values. Similar to the previous example, computations based only on changes of porosity underestimate the permeability loss exhibiting larger discrepancies from the data towards the end of the test, i.e. when larger pressures promote a more intense growth of the surface area.\\
The calibrated model has also been used to predict the mechanical response and permeability reduction of Glass Beads subjected to hydrostatic loading ($r$= 1.0). Figures \ref{figure5}d-\ref{figure5}f illustrate the computed compression response, grain size distribution and permeability reduction by comparing them against the experimental data. The results show that the predictions have the same qualitative trends exhibited by the experimental data and despite the limited dataset used for parameter calibration they provide a satisfactory estimate of the permeability at the end of the compression stage  \cite{Nguyen_2011}. 
\subsection{Cemented granular rocks}
\begin{figure}[!ht]
\centering
\includegraphics[width=13.0cm]{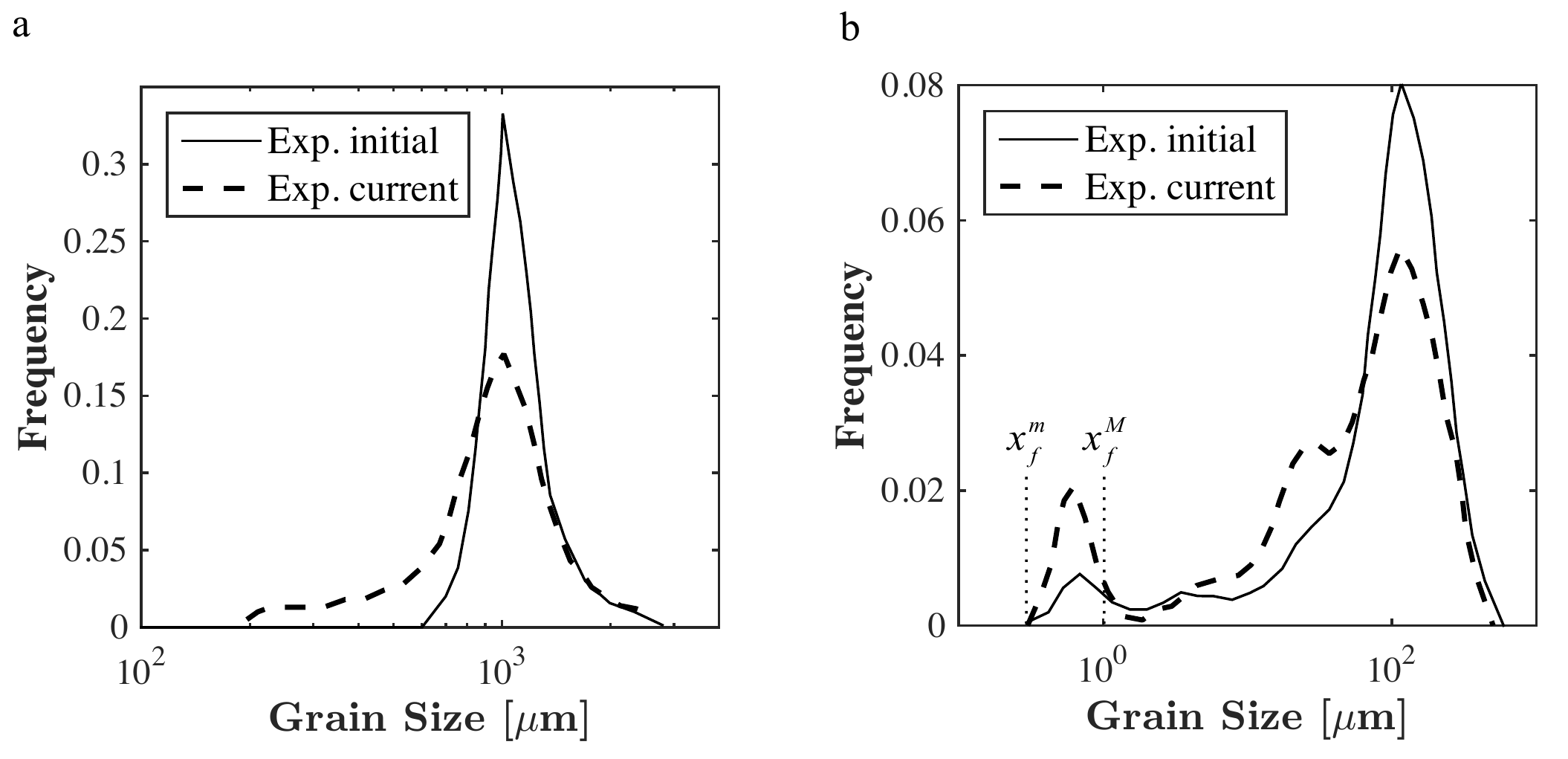}
\caption{Grain size distributions of (a) Glass Beads, (b) Otter Sherwood sandstone, before and after isotropic compression tests [data re-plotted after \cite{Nguyen_2014}].}\label{figure6}
\end{figure}

Compression processes may alter the microstructure of a granular solid in very different ways depending on whether or not a cementing agent is present. An example of this concept is illustrated in Figure \ref{figure6}, where the gsd of a cohesionless granular material before and after loading is compared with the initial and final gsd curves of a weakly cemented sandstone. It is readily apparent that, while for the granular material the increase in particle dispersion is accompanied by the preservation of unimodality, in the sandstone the compression process exacerbates the bimodality of the initial particle distribution. The second peak emerging at the lower bound of the distribution, indicates the generation of small grains that can be related to the disaggregation of the binding agent and the creation of fine particulate (hereafter more simply referred to as "fines"), (Figure \ref{figure2}). Since these fines participate to the evolution of the surface area and can have important implications on the hydraulic conductivity, (i.e. they clog the pores and increase the solid surface interacting with the fluid), they have to be taken into account in the calculation of the grain size distribution and consequently of $D_H$. For the sake of simplicity, hereafter it will be hypothesized that all the fine particulates emerging from the disaggregation of the cement hinder the flow of water, thus further exacerbating the permeability loss. This hypothesis will provide a lower bound to the computed values of permeability, which will be compared against the scenarios in which the effect of the cement fines is entirely neglected.\\
Here, to estimate the accumulation of fines in relation with the damage of the cement matrix, the data illustrated in Figure \ref{figure6} are used as a basis to define a simple geometric model for the fines resulting from cement disaggregation. By assuming to have an ultimate cumulative fine size distribution $F_u^f(x_f)$ with minimum size $x_f^m$ and maximum size $x_f^M$, the current cumulative fine size distribution can be written as a linear function of the Damage index: $F^f(x_f,D)=DF_u^f(x_f)$. This relation implies that no fines are present when the cement matrix is undamaged, while the fine distribution achieves ultimate condition when the cement is locally fully damaged. As a result, the general expression for the grain size distribution considering both the skeletal particles and the fines generated by the disaggregating cement can be obtained as:
\begin{equation}\label{general_gsd}
F(d,B,D) = \frac{\varphi^g}{\varphi^g+\varphi^f}F^g(d,B)+\frac{\varphi^f}{\varphi^g+\varphi^f}F^f(x_f,D)
\end{equation}
where $\varphi^f$ is the current volume fraction of the fines calculated as $\varphi^f=D\varphi^c$. By analogy to Equation (\ref{harmonic_mean1}), the general expression for $D_H$ is given by:
\begin{equation}\label{general_DH}
\frac{1}{D_H(B,D)}=  \frac{\varphi^g}{\varphi^g+\varphi^f}((1-B)\frac{1}{D_{H0}}+B\frac{1}{D_{Hu}})+ \frac{\varphi^f}{\varphi^g+\varphi^f}\frac{D}{D^f_{Hu}}
\end{equation}
which underpins an evolution of the material surface area as a function of both Breakage and Damage. From Equation (\ref{general_DH}, it is readily apparent that by setting $\varphi^f=0$, (\ref{general_gsd}) and (\ref{general_DH}) reduce to (\ref{current_cpdf}) and (\ref{harmonic_mean1}) respectively, thus recovering the relations valid for cohesionless granular systems as a particular case.
\subsubsection{Otter Sherwood sandstone}
\begin{figure}[!ht]
\centering
\includegraphics[width=11.0cm]{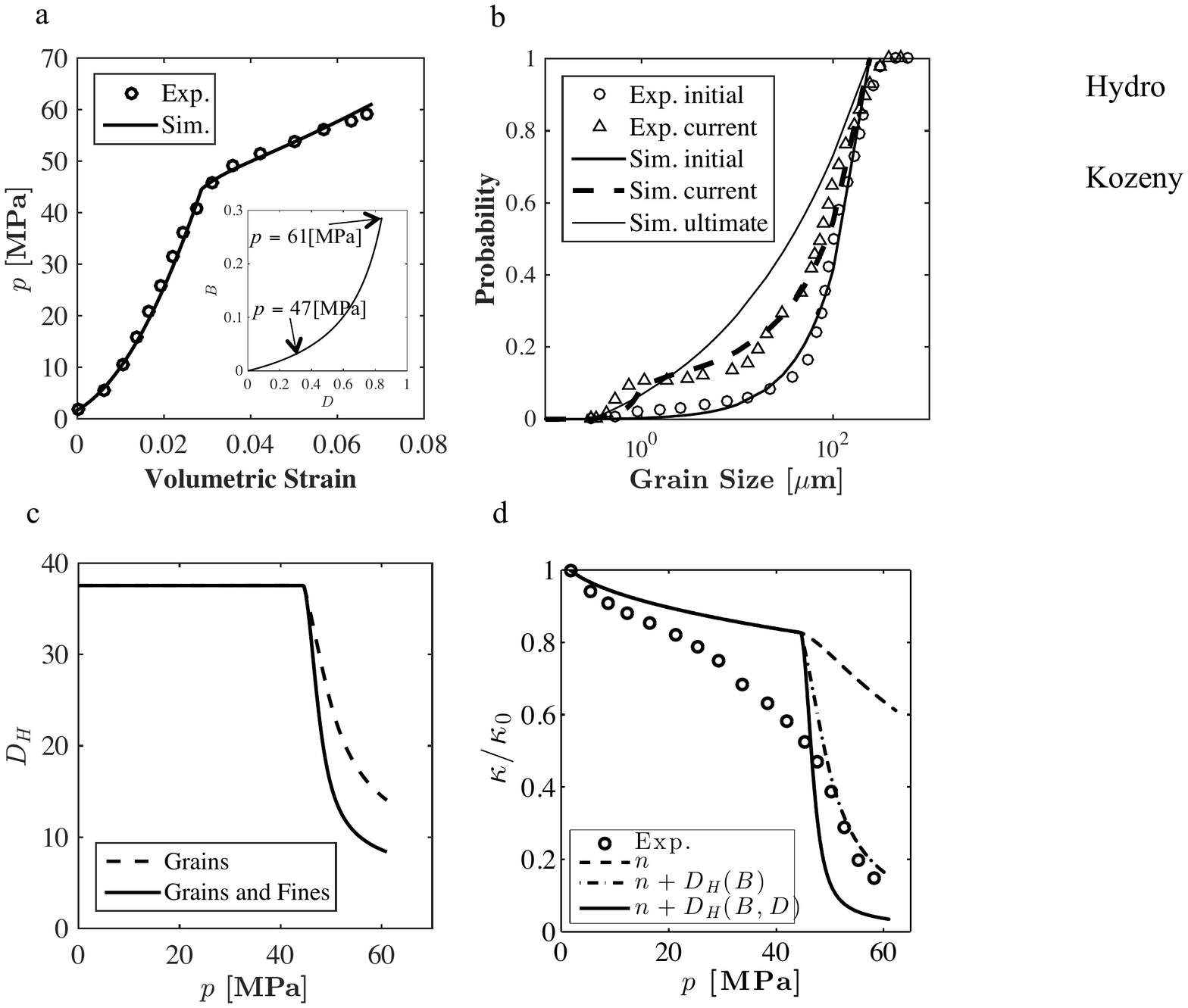}
\caption{(a) Mechanical response, (b) Grain size distribution evolution, (c) Harmonic mean grain size $D_{H}$ and (d) Permeability approximation for Otter Sherwood sandstone under isotropic compression test [data re-plotted after \cite{Nguyen_2014}].}\label{figure7}
\end{figure}

\begin{figure}[!ht]
\centering
\includegraphics[width=6.0cm]{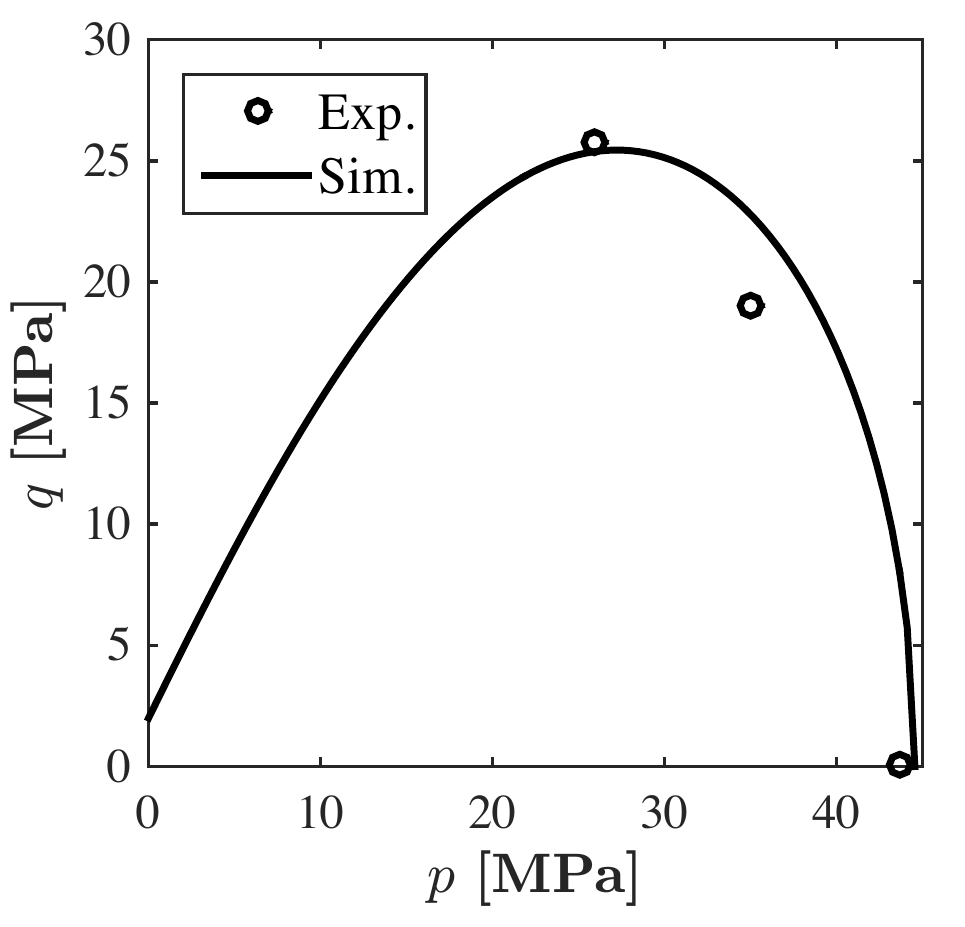}
\caption{Yield surface of the Breakage-Damage model for Otter Sherwood sandstone represented in the triaxial stress space [data re-plotted after \cite{Nguyen_2014}].}\label{figure10}
\end{figure}

\begin{figure}[!ht]
\centering
\includegraphics[width=11.0cm]{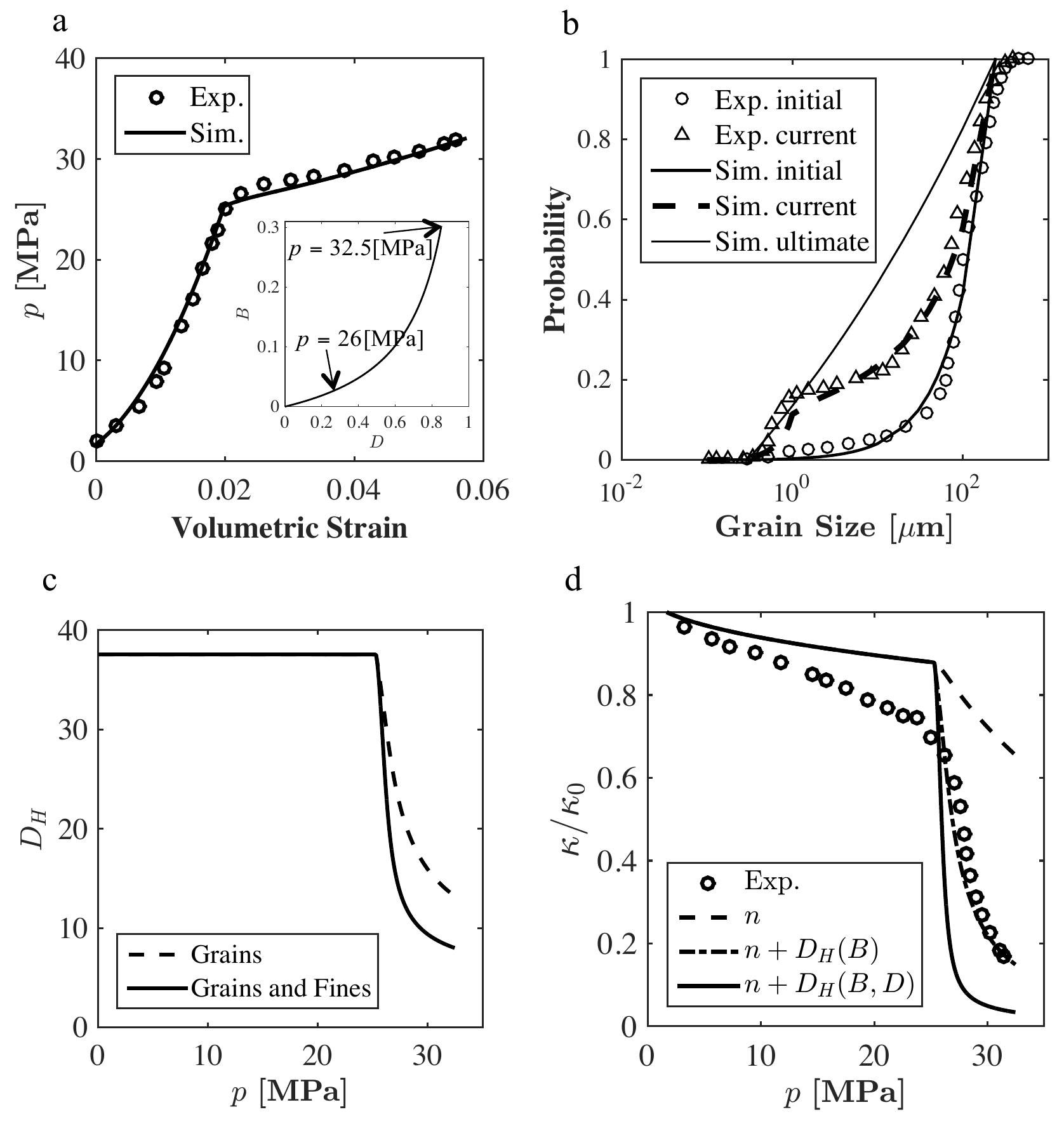}
\caption{(a) Mechanical response, (b) Grain size distribution evolution, (c) Harmonic mean grain size $D_{H}$ and (d) Permeability approximation for Otter Sherwood sandstone under radial path loading of $r$= 0.4 [data re-plotted after \cite{Nguyen_2014}].}\label{figure8}
\end{figure}

\begin{figure}[!ht]
\centering
\includegraphics[width=11.0cm]{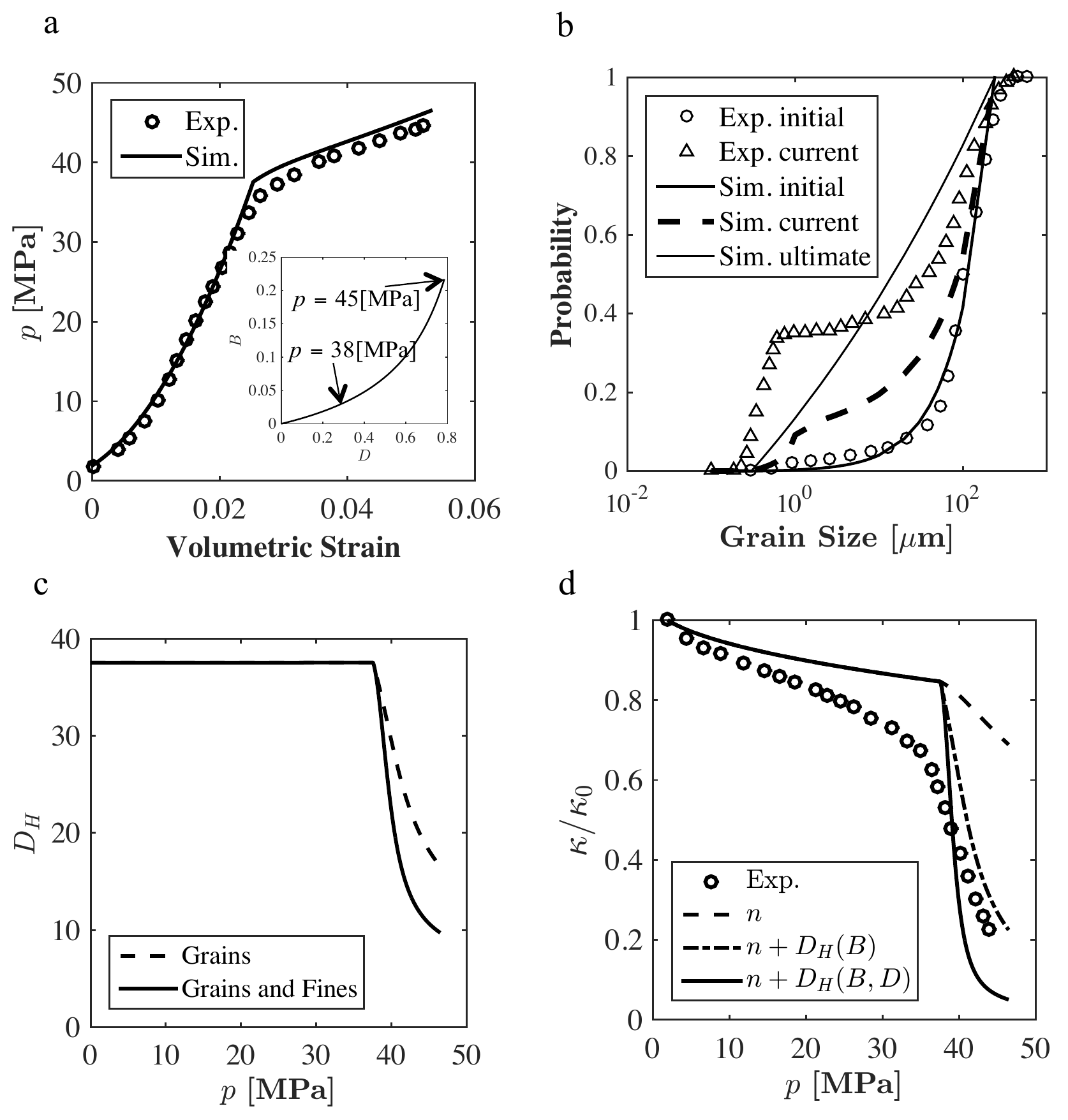}
\caption{(a) Mechanical response, (b) Grain size distribution evolution, (c) Harmonic mean grain size $D_{H}$ and (d) Permeability approximation for Otter Sherwood sandstone under radial path loading of $r$= 0.6 [data re-plotted after \cite{Nguyen_2014}].}\label{figure9}
\end{figure}

Otter Sherwood sandstone is a weakly consolidated rock which has been characterized through hydro-mechanical laboratory tests by \cite{Nguyen_2014}. Since this material is weakly cemented, a cement volume fraction $\varphi^c= 10\%$ has been used, in agreement with the range reported in the literature for rocks from the same formation \cite{Holloway_1989}. The isotropic compression test is the first loading path used for model calibration, and it has allowed to obtain the model response shown in Figure \ref{figure7}a (calibrated parameters: $\overline{K}^g$=$\overline{K}^c$= 11900, $E_{BC}$=$E_{DC}$= 0.3 MPa, $\omega_B$= 74$^{\circ}$).\\
Figure \ref{figure7}b compares the computed grain size distribution against the experimental data. Assuming $d_m$= 1.0 $ \text{$\mu$m}$ and $d_M$= 240 $ \text{$\mu$m}$ for both the initial and ultimate cumulative gsd functions, $F_0(d)$ is approximated by a power law distribution with the power law coefficient of 2, while a fractal distribution with the fractal dimension $\alpha$ of 2.7 is used for ultimate grading $F_u(d)$ and $\vartheta^g$= 0.52 is calculated from $F_0(d)$ and $F_u(d)$. The current grain size distribution is derived by assuming to have a uniform distribution for the generated fines with $x_f^m$= 0.1 $ \text{$\mu$m}$ and $x_f^M$= 1.0 $ \text{$\mu$m}$ and using the final Breakage value at the end of the simulation ($B$=0.29) and the final Damage value of $D$= 0.85. It can be seen that by considering the generation of fines resulting from the damage of the cement matrix, the simulated gsd is very close to that reported in the experiment.\\
For permeability predictions, $D_H$ has been calculated for two scenarios: 1) by neglecting the cement fines and considering solely the skeletal grains and 2) by including the effect of the cement fines in the computation of $D_H$. Both cases are plotted in Figure \ref{figure7}c and then used in the Kozeny equation to estimate the variation of the rock permeability along with the case of neglecting the surface area evolution (Figure \ref{figure7}d). It is possible to notice that, despite the tendency of the model to underestimate the loss of permeability prior to yielding, the model captures the first-order features of the measured response. In addition, the permeability approximation using the $D_H$ for only grains and the combination of grains and fines are very close to each other as well as the values of permeability measured in the experiment and provide the upper and lower bound scenarios for permeability prediction depending on the amount of the fines participated in the permeability reduction processes. For this specific cemented granular rock, the prediction obtained by taking grain breakage into account and excluding the cement fines from the analysis is found to be the one closest to the experimental data, thus suggesting that the contribution of the fines to the overall permeability reduction tends to be much smaller than that associated with the crushing of the grain skeleton. From the same figure it is also readily apparent that the values of permeability computed by the models accounting for surface area growth are significantly more accurate than those based only on porosity changes, from which a considerable underestimation of the permeability losses emerges at all the stages of the simulation.\\
While the hydrostatic test is used to define the compressive properties and the proportion between breakage and damage dissipation, information about the pressure dependence of yielding is used to define the properties associated with the shearing behavior, such as shear modulus, friction angle and cohesion (i.e., all properties that affect the size and shape of the yield envelope predicted by the model). For this purpose, the yield function (Equation (\ref{yield})) is plotted in Figure \ref{figure10} against the yield points obtained from the experiments and the model simulations for three loading paths, $r$= 1.0, 0.4 and 0.6 tests and the remaining parameters are calibrated accordingly ($G^g=G^c$= 1400 MPa, $c$= 2 MPa and $M$= 1.45).\\
After having all the model parameters calibrated, the radial path loading with $r$= 0.4 was modeled and the corresponding results are plotted in Figure \ref{figure8}a. A very good agreement can be seen between the model mechanical response and the experiments.\\
By assuming the same initial and ultimate grain size distributions and ultimate fine size distributions, the grain size distribution computed at the end of the experiment is reported in Figure \ref{figure8}b. Again two  values of $D_H$ are calculated, as reported in Figure \ref{figure8}c and then used in the analyses in Figure \ref{figure8}d. The computed trends are similar to those previously shown with reference to isotropic compression, having once again that the permeability reduction resulting from the predicted degree of grain breakage again proves to be the most accurate. By contrast, the estimates based only on porosity changes are found again to severely underestimate the permeability loss, while those accounting for the potential effect of cement fines tend to exacerbate the computed rate of permeability reduction.\\
The calibrated model can eventually be used to predict the mechanical response (Figure \ref{figure9}a), the grain size distribution at the end of the experiment (Figure \ref{figure9}b), the evolution of $D_H$ (Figure \ref{figure9}c) and of the permeability (Figure \ref{figure9}d) for the radial path loading of $r$= 0.6. By comparing the model prediction with the experimental data, a good agreement is observed for the mechanical response and the permeability reduction. A significant mismatch between data and model predictions can  instead be observed in terms of final gsd for the test performed with $r$= 0.6. In the latter case, a much larger generation of fines was measured compared to the tests performed at $r$= 1.0 and $r$ = 0.4. Although this result can be affected by inevitable experimental errors involved with the measurement, it also suggests the presence of inherent differences between the three samples, with the sample subjected to $r$= 0.6 possibly characterized by a cementing phase with different microstructural properties.\\
\section{Conclusions}\label{conclusion}
The analysis of permeability changes in granular rocks subjected to high compressive stresses has to account for significant microstructural alterations such as grain crushing, cement damage and pore collapse. To address this challenge, this paper uses the Breakage Mechanics framework and links an inelastic constitutive law for crushable granular solids to a permeability model for granular rocks. The proposed model tracks the changes in porosity and surface area by expressing them as a function of internal variables associated with the evolution of the grain size distribution and the cement size distribution of the rock. This choice enables the coupling of the deformation response with permeability evolution laws, and has been here exploited by calibrating the model parameters for two different granular materials (i.e., Glass Beads and Ottawa sand) and a weakly cemented granular rock (Otter Sherwood sandstone). The former set of analyses are aimed to study a simple proxy of unconsolidated rock, as well as to validate the use of the Kozeny equation in combination with a Breakage law for cohesionless materials. The latter rock is instead used with the goal to assess the role of cement disaggregation on the growth of surface area and the consequent reduction of the rock permeability. It is shown that for such materials, the mechanical response, grain size distribution (gsd) evolution and permeability reduction can be predicted satisfactorily for different loading paths by linking Breakage Mechanics formulation to a standard permeability equation for granular media, i.e., the Kozeny equation, which here has been expressed as a function of the Breakage internal variables through harmonic mean grain size. In particular, it has been shown that the most accurate predictions of loss of hydraulic conductivity are obtained by the permeability model relying on both porosity changes and surface area growth due to grain crushing. By contrast, it has been found that permeability models relying only on porosity changes tend to severely overestimate the value of rock permeability (thus providing an upper bound), while those accounting for the effect of surface area change due to cement disaggregation underestimate its values (thus representing a lower bound for the expected permeability of a comminuted rock). The performance of the proposed hydro-mechanical models suggests that this approach may provide enhanced capabilities for the simulation of evolving porosity and permeability in reservoir rocks, thus representing a convenient tool to connect the large-scale implications of fluid depletion to their inherent microscopic causes. As a result, this work motivates future analyses that can further expand the scope of microscopic processes taken into account for the computation of permeability reduction, such as the rupture of asperities and the generation of fines prior to macroscopic yielding, the heterogeneity of the mineral content in the cement and skeleton phases and the possible emergence of heterogeneous deformation processes in the form of compaction and/or shear bands.
%
\section*{Appendix}\label{appendix}
The volumetric and shear part of the reference energy in Breakage Mechanics models can be assumed to be linear as:
\begin{equation}\label{flow_rules}
\begin{split}
&\Psi_r^{v,\ell}=\frac{1}{2} K^\ell (\varepsilon_v^e)^2\\ 
&\Psi_r^{s,\ell}=\frac{3}{2} G^\ell (\varepsilon_s^e)^2\\
\end{split}
\end{equation}
where $\ell=g,c$ represents the grain and the cement phase, $K^\ell$ is the bulk stiffness of the grains or the cement bond and $G^\ell$ is the shear stiffness of these attributes. $\varepsilon_v^e$ is the elastic volumetric strain and $\varepsilon_s^e$ is the elastic shear strain.\\
On the other hand, these reference energy functions could also be considered nonlinear or pressure-dependent in the form of:
\begin{equation}\label{flow_rules}
\begin{split}
&\Psi_r^{v,\ell}=\frac{2}{3} p_r \frac{(A^\ell)^3}{\overline{K}^\ell}\\ 
&\Psi_r^{s,\ell}=\frac{3}{2} p_r \overline{G}^\ell A^\ell (\varepsilon_s^e)^2\\
\end{split}
\end{equation}
$p_r=$1 kPa is the reference pressure, $\overline{K}^\ell$ is the non-dimensional material parameter representing the bulk stiffness for grains or cement bond, $\overline{G}^\ell$ is the non-dimensional material parameter representing the shear stiffness for grains or cement bond and $A^\ell=\overline{K}^\ell \varepsilon_v^e/2+1$.\\
Depending on the material's elastic response, either linear, nonlinear or even hybrid forms (i.e. nonlinear function for the volumetric part and linear function for the shear part and vice versa) of the reference energy can be selected. In this paper, nonlinear elasticity had been used for cohesionless materials for both volumetric and shear reference energy functions while a combination of nonlinear volumetric and linear shear functions has been used for the cemented granular rocks.\\
In addition, the flow rules for this model derived from the dissipation potential \cite{tengattini_2014} are listed in Equation (\ref{flow_rules}).
\begin{equation}\label{flow_rules}
\begin{split}
&\dot{B}=2\dot{\lambda}\frac{(1-B)^2E_B\cos(\omega_B)}{E_B E_{BC}}\\ 
&\dot{D}=2\dot{\lambda}\frac{(1-D)^2E_D}{E_D E_{DC}}\\
&\dot{\varepsilon_v}^p=2\dot{\lambda}\frac{(1-B)^2E_B\sin(\omega_B)}{pE_{BC}}\\
&\dot{\varepsilon_s}^p=2\dot{\lambda}\frac{q}{Mp+c(1-D)^2}
\end{split}
\end{equation}
where $\dot{\lambda}$ is a non-negative plastic multiplier, $\varepsilon_v^p$ and $\varepsilon_s^p$ are the plastic volumetric and shear strains, respectively, and $\omega_B$ is a model parameter coupling grain crushing and the volumetric plastic dissipation associated with the reorganization of crushed fragments \cite{nguyen_2012_1}.
\section*{Acknowledgments} 
This work was supported by Grant No. CMMI-1351534 awarded by the U.S. National Science Foundation.
\section*{References}
\bibliography{ALLS}
%
\end{document}